\documentclass[twocolumn]{aastex631}
\usepackage{CJK}
\usepackage[T1]{fontenc} 
\usepackage[utf8]{inputenc}

\usepackage{amsfonts}
\usepackage{amsmath}
\usepackage{mathrsfs}
\usepackage{epsfig}
\usepackage{subfigure}
\usepackage{multirow}
\usepackage{threeparttable}

\newcommand\diff{\,\mathrm{d}}

\def\BJD{\,\mathrm{BJD}}

\def\Porb{\,P_\mathrm{orb}}
\def\Teff{\,T_{\mathrm{eff}}}
\def\dPdt{\,(1/P_0)(\diff P_0/\diff t)}
\def\yr{\,\mathrm{yr}}
\def\logg{\,\log g}
\def\cd{\mathrm{c\ d}^{-1}}
\def\Msun{\,\mathrm{M}_{\odot}}
\def\raa{Res.\ Astron.\ Astrophys.\ }

\shorttitle{CY Aqr's Twin Companions}
\shortauthors{Lian, Niu \& Xue}

\graphicspath{{./}{figures/}}

\begin{document}
\begin{CJK*}{UTF8}{gbsn}

\title{CY Aquarii: A Triple System with Twin Companions}

\author{Ming Lian (连茗)}
\affil{Institute of Theoretical Physics, Shanxi University, Taiyuan 030006, China}
\affil{State Key Laboratory of Quantum Optics Technologies and Devices, Shanxi University, Taiyuan 030006, China}

\author[0000-0001-5232-9500]{Jia-Shu Niu (牛家树)}
\affil{Institute of Theoretical Physics, Shanxi University, Taiyuan 030006, China}
\affil{State Key Laboratory of Quantum Optics Technologies and Devices, Shanxi University, Taiyuan 030006, China}
\affil{Collaborative Innovation Center of Extreme Optics, Shanxi University, Taiyuan 030006, China}
\correspondingauthor{Jia-Shu Niu}
\email{jsniu@sxu.edu.cn}

\author[0000-0001-6027-4562]{Hui-Fang Xue (薛会芳)}
\affil{Department of Physics, Taiyuan Normal University, Jinzhong 030619, China}
\affil{Institute of Computational and Applied Physics, Taiyuan Normal University, Jinzhong 030619, China}
\affil{Shanxi Key Laboratory for Intelligent Optimization Computing and Blockchain Technology, Jinzhong 030619, China}

\begin{abstract}
In this study, we conducted a comprehensive analysis of the SX Phoenicis (SX Phe) type star CY Aquarii (CY Aqr). Our investigation included a detailed $O-C$ analysis based on a 90-year observational dataset, augmented by 1,367 newly determined times of maximum light. The $O-C$ diagram reveals that (i) the primary star of CY Aqr exhibits a linear period variation rate of $ (1/P_0)(\diff P/dt) = (2.132 \pm 0.002) \times 10^{-8} \yr^{-1}$ for its dominant pulsation mode; (ii) the primary star is disturbed by two companions and part of a triple system; (iii) Companion A has an orbital period of approximately 60.2 years and Companion B has an orbital period of approximately 50.8 years. It is highly probable that both Companion A and B are white dwarfs, with Companion A's elliptical orbit displaying an eccentricity of $e = 0.139 \pm 0.002$, which is the lowest confirmed value in similar binary and triple systems to date.
Most notably, Companion A and B have masses that are identical within the uncertainties, with a mass ratio exceeding 0.99. Whether this is considered a coincidental event or the result of an underlying mechanism, CY Aqr is an exceptionally rare case that broadens our understanding of multiple star systems and offers a unique opportunity to delve into the enigmatic evolutionary histories of such configurations. Further intriguing characteristics of this system warrant investigation in future studies, based on additional observational data.
\end{abstract}

%\keywords{Pulsating variable stars --- stars: variables: SX Phoenicis variable stars --- Asteroseismology}

\section{Introduction}
\label{sec:intro}

SX Phoenicis (SX Phe) variables, characterized as short-period pulsation stars residing within the classical Cepheid instability strip, are distinguished by their low metallicity and high spatial velocity \citep{McNamara1995}. These high-amplitude $\delta$ Scuti stars (HADS) of Population II display variability, marked by either single or double radial modes and significant pulsation amplitudes \citep{Yang2012,Daszy2020,Niu2023}, akin to those observed in HADS \citep{Niu2017,Xue2018,Bowman2021,Niu2022,Xue2022,Xue2023,Daszy2023,Daszy2024}. The SX Phe variables, with masses ranging from 1.0 to 1.2 $M_{\odot}$ and ages spanning approximately 2 to 5 Gyr \citep{Nemec1990}, present an enigmatic origin and evolutionary mechanism, with hypotheses suggesting a possible genesis through the merger of close binary stars \citep{Breger2000}. Consequently, the examination of SX Phe stars within binary and multiple star systems is crucial for unraveling their formation and evolutionary narratives.

CY Aquarii (CY Aqr, $\langle V \rangle = 11.04$ mag, spectral type: A2-A8, $\mathrm{[Fe/H]} = -1.5$ \citep{McNamara1996}) represents an SX Phe variable star in the post-main-sequence phase of evolution \citep{Andreasen1983}, initially discovered by Hoffmeister \citep{Jensch1934} and subject to extensive, long-term observations (for example, see \citet{Struve1949, Hardie1961, Nather1972}). While confirmed as a radial single-mode pulsating star \citep{Xue2023}, the specific mode of pulsation, whether fundamental or overtone, remains uncertain \citep{McNamara1995,pena1999}.

Regarding the period variation of CY Aqr, \citet{Hardie1961} noted a subsequent decrease in the star's period during the 1950s. \citet{percy1975} later concluded that the period was stable between 1934 and 1951, with a significant alteration in 1951. As research into the period variation persisted, multiple theories emerged to account for this phenomenon \citep{mahdy1980,kamper1985,powell1995}. With the aggregation of photometric data, the prevailing explanation for CY Aqr's period variation shifted to the light-travel time effect (LTTE) within a binary star system \citep{coates1994,fu2003,Derekas2009}. Nevertheless, periodic structures in the $O - C$ diagram residuals demand attention \citep{fu2003}. \citet{fang2016} posited that the target star resides within a triple system, with the orbital periods of the two companion stars A and B determined to be 54.2 years and 47.3 years, respectively.

On one front, photometric observations of CY Aqr have been ongoing, particularly the high-precision, continuous data from the Transiting Exoplanet Survey Satellite (TESS)\footnote{\url{https://tess.mit.edu/science/}}, which will not only elucidate further pulsation properties of the primary star but also yield more precise times of maximum light. Concurrently, contemporary numerical frameworks enable a more rational and precise determination of the system's orbital parameters \citep{Xue2020}. Thus, it is apt to revisit this system.

\section{Methods and Results} 
\label{sec:methods}

\subsection{Pulsation Analysis}
\begin{figure}[!htbp]
	\centering
	\includegraphics[width=0.49\textwidth]{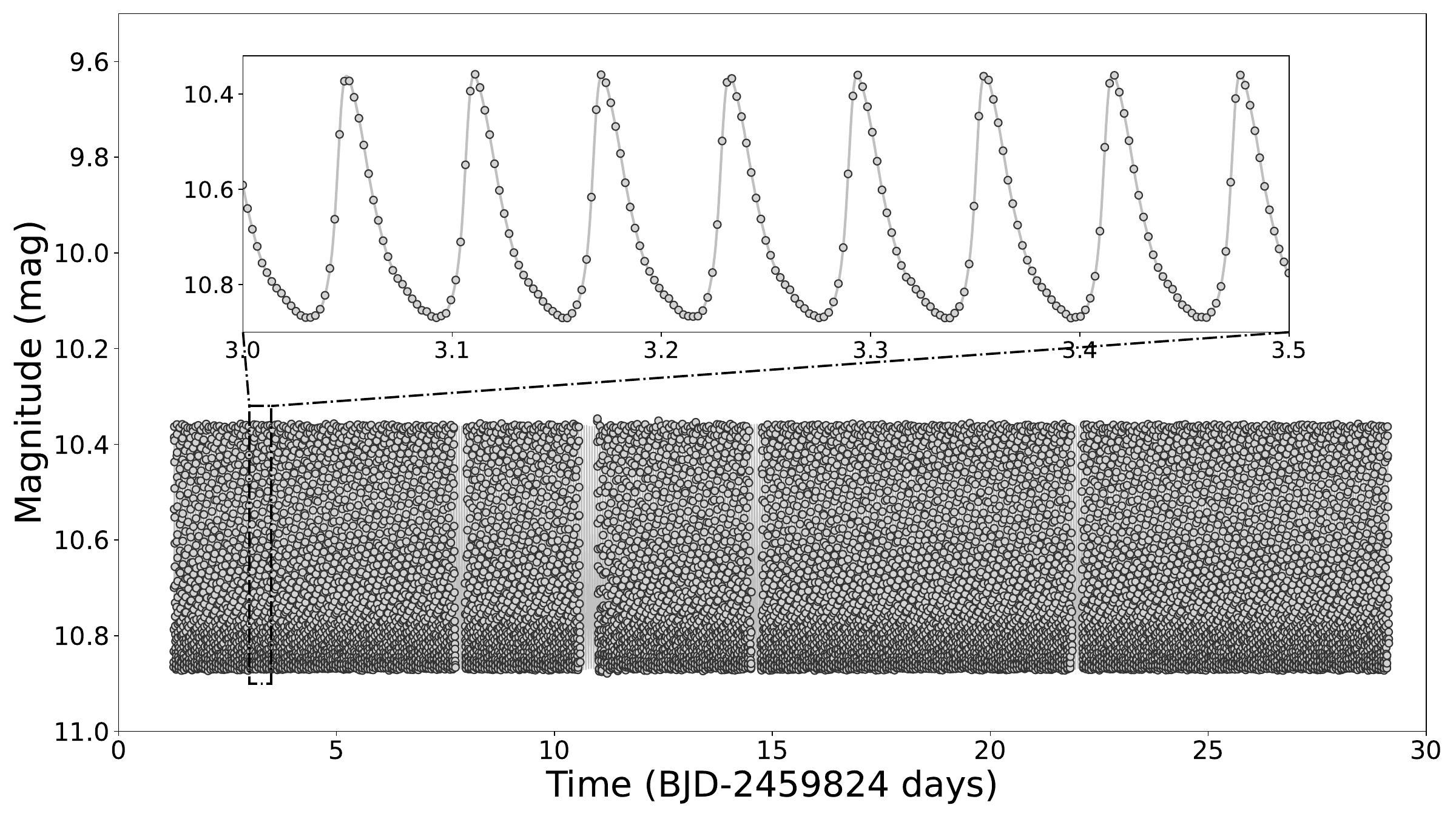}
	\caption{Light curve of CY Aqr (TESS Sector 56). The data points are represented by dots, and the gray line indicates the curve fitted using 22 extracted frequencies. The top panel displays a zoomed-in area covering $0.5\ \mathrm{d}$.}
	\label{fig:lc}
\end{figure}

The Transiting Exoplanet Survey Satellite (TESS) is a NASA mission dedicated to the detection of transiting exoplanets through an all-sky survey. CY Aqr was observed by TESS at a 120-s cadence during Sector 42 (August 2021) and at a 200-s cadence during Sectors 56 (September 2022) and 70 (September 2023). The data from these observations are accessible from the Mikulski Archive for Space Telescopes (MAST)\footnote{\citet{CYAqr_data}, \url{http://archive.stsci.edu/doi/resolve/resolve.html?doi=10. 17909/3gqn- pz22}}. The Presearch Data Conditioning Simple Aperture Photometry (PDCSAP) flux was converted to magnitudes utilizing the TESS magnitude system \citep{smith2012,stumpe2012,stumpe2014,Stassun2019a}. The light curves of CY Aqr from TESS Sector 56 are depicted in Figure \ref{fig:lc}.

The prewhitening process of the light curves (TESS Sector 56) was executed using the Period04 software \citep{Lenz2005}. The light curves were modeled with the following formula:
\begin{equation}
	 m = m_0 + \sum a_i \sin[2\pi(f_i t + \phi_i)] ,
\end{equation}
where $m_0$ is the zero-point magnitude, $a_i$ is the amplitude, $f_i$ is the frequency, and $\phi_i$ is the corresponding phase. We applied a signal-to-noise ratio (S/N) threshold of greater than 5.6 as the criterion for frequency significance \citep{Niu2022,Niu2023}.

\begin{figure*}[!htbp]
	\centering
		\includegraphics[width=0.49\textwidth]{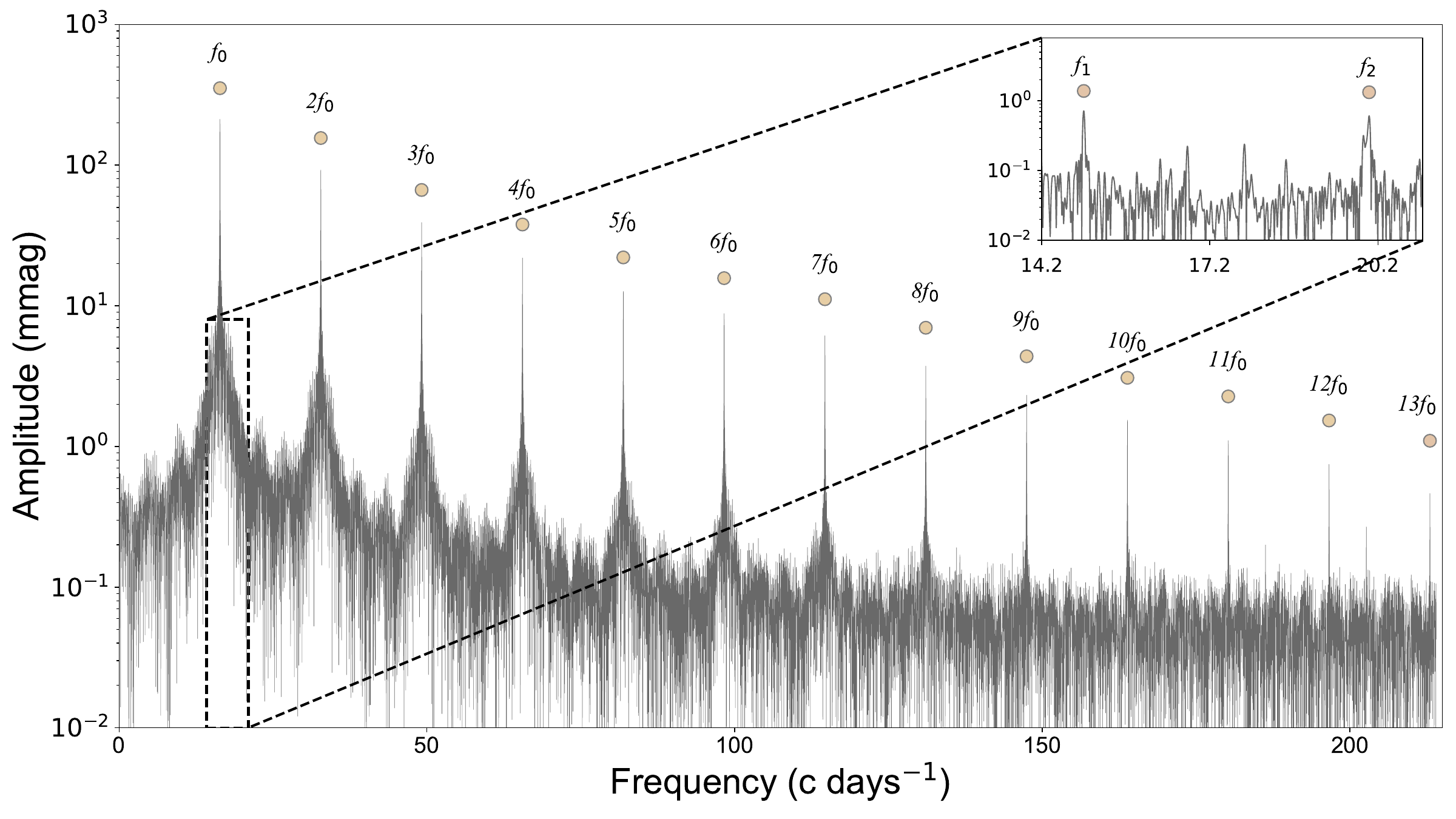}
		\includegraphics[width=0.49\textwidth]{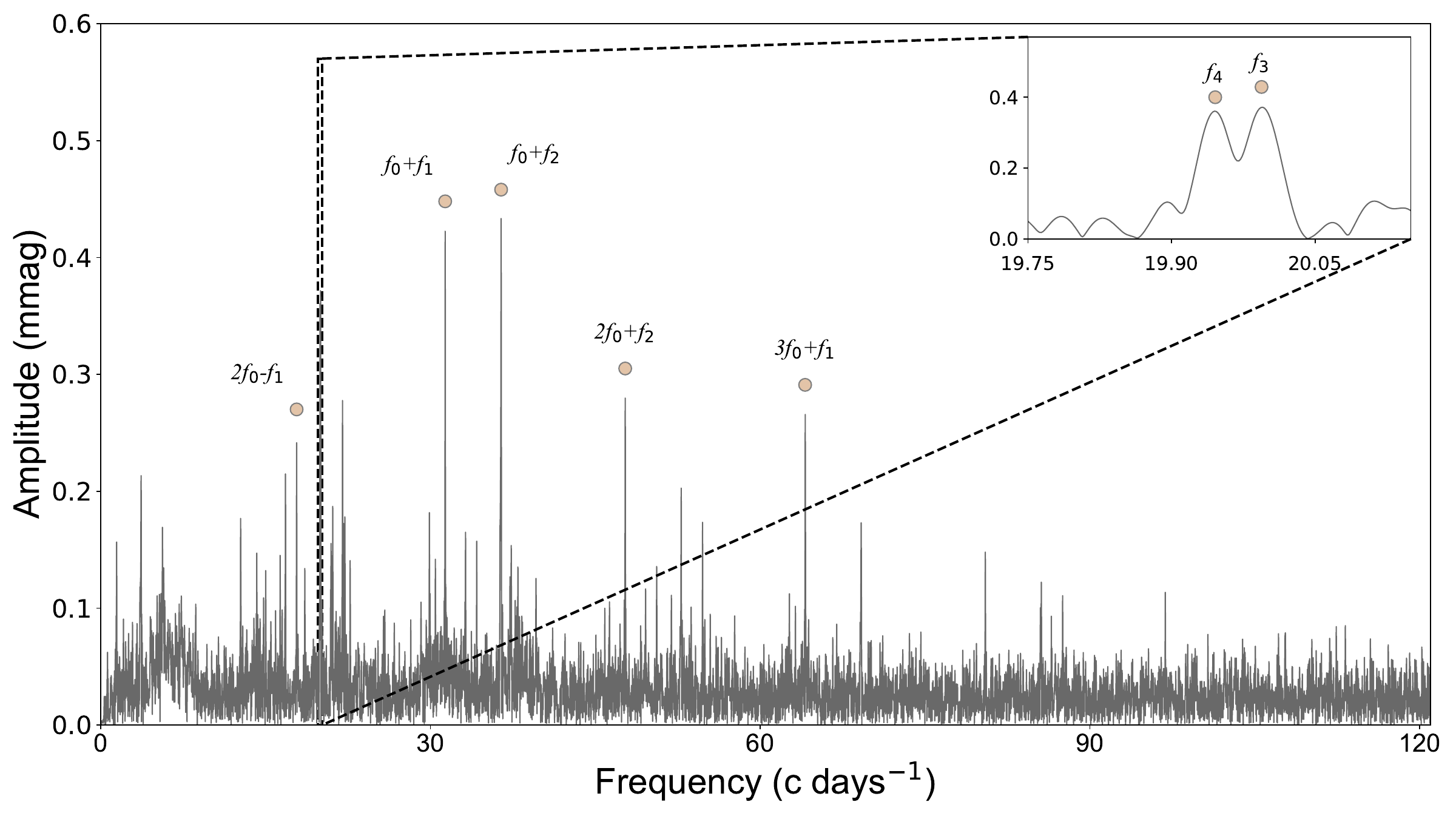}
	\caption{Frequency spectra of the light curve of CY Aqr. The left panel displays the range $0-215 \ \cd$, with $f_0$ and its harmonics, $f_1$, and $f_2$ represented. The right panel shows the range $0-120 \ \cd$, with $f_3$ and $f_4$ represented after the removal of $f_0$ and its harmonics, $f_1$, and $f_2$.}
	\label{fig:spectra}
\end{figure*}

After the removal of alias frequencies, 22 statistically significant frequencies were identified, comprising 5 independent frequencies and 17 harmonics/combinations. For further details, refer to Figure \ref{fig:spectra} and Table \ref{tab:freqs}.

\begin{table*}[htp]
   \centering
		\caption{Multifrequency Solution of the Light Curves of CY Aqr from TESS (Sector 56).\label{tab:freqs}}	
			\begin{tabular}{ccccccc}
				\hline
				\hline
				NO. & Marks & $f \ (\cd)$ & $\sigma_f \ (\cd)$  & $a \ (\mathrm{mmag})$ & $\sigma_a \ (\mathrm{mmag})$ & S/N \\
				\hline
				F1    & $f_0$   & 16.383116  & 0.000002  & 212.01  & 0.02 
				  & 18.9  \\
				F2    & 2$f_0$   & 32.766246  & 0.000005  & 92.87  & 0.02  & 19.0  \\
				F3    & 3$f_0$   & 49.14937  & 0.00001  & 39.17  & 0.02  & 18.9  \\
				F4    & 4$f_0$   & 65.53250  & 0.00002  & 22.09  & 0.02  & 18.9  \\
				F5    & 5$f_0$   & 81.91564  & 0.00004  & 12.78  & 0.02  & 18.9  \\
				F6    & 6$f_0$   & 98.29865  & 0.00005  & 8.93  & 0.02  & 18.8  \\
				F7    & 7$f_0$   & 114.68173  & 0.00008  & 6.13  & 0.02  & 18.9  \\
				F8    & 8$f_0$   & 131.0650  & 0.0001  & 3.73  & 0.02  & 18.2  \\
				F9    & 9$f_0$   & 147.4480  & 0.0002  & 2.31  & 0.02  & 17.8  \\
				F10   & 10$f_0$  & 163.8308  & 0.0003  & 1.57  & 0.02  & 17.4  \\
				F11   & 11$f_0$  & 180.2145  & 0.0005  & 1.08  & 0.02  & 16.9  \\
				F12   & $f_1$    & 14.9502  & 0.0007  & 0.72  & 0.02  & 10.6  \\
				F13   & 12$f_0$  & 196.5967  & 0.0007  & 0.71  & 0.02  & 14.9  \\
				F14   & $f_2$    & 20.0421  & 0.0007  & 0.71  & 0.02  & 8.7  \\
				F15   & 13$f_0$  & 212.980  & 0.001  & 0.49  & 0.02  & 13.0  \\
				F16   & $f_0 + f_2$ & 36.425  & 0.001  & 0.43  & 0.02  & 7.2  \\
				F17   & $f_0 + f_1$ & 31.334  & 0.001  & 0.42  & 0.02  & 7.5  \\
				F18   & $f_3$    & 19.994  & 0.001  & 0.40  & 0.02  & 6.9  \\
				F19   & $f_4$    & 19.946  & 0.001  & 0.37  & 0.02  & 8.0  \\
				F20   & 2$f_0 + f_1$ & 47.717  & 0.002  & 0.28  & 0.02  & 7.4  \\
				F21   & 3$f_0 + f_1$ & 64.100  & 0.002  & 0.27  & 0.02  & 6.5  \\
				F22   & 2$f_0 - f_1$ & 17.816  & 0.002  & 0.24 & 0.02  & 7.2  \\
				\hline
				\end{tabular}
\end{table*}

\subsection{Pulsation Mode Identification}
\label{subsec:ModeIden}

The pulsation constant $Q$ is defined by the formula:
\begin{equation}
	Q = P \sqrt\frac{\bar\rho}{\bar\rho_\odot} ,
\end{equation}
where $P$ represents the pulsation period, $\bar\rho$ is the mean density of the star, and $\bar\rho_\odot$ is the mean density of the Sun. Given the difficulty in directly detecting the mean density of stars, their pulsation constants $Q$ are derived using the following relationship \citep{Breger1990}:
\begin{equation}
	\log Q = \log P + 0.5 \log g + 0.1 M_\mathrm{bol} + \log \Teff - 6.456.
\end{equation}

As the pulsation constant $Q$ is sensitive to surface gravity and effective temperature \citep{Poro2024}, we employed two models—the TESS Input Catalog and Gaia DR3—as illustrated in Table \ref{tab:pulsation_constant}—to ascertain $Q$ \citep{Stassun2019a, Gaia2021}.

The absolute magnitude $M_\mathrm{V}$ is calculated using the equation:
\begin{equation}
	M_{V} = V - 5 \log {d} + 5 - A_{V},\label{Mv}
\end{equation}
and the absolute thermal magnitude $M_\mathrm{bol}$ is determined according to \citet{Petersen1999}:
\begin{equation}
	M_\mathrm{bol}=M_{V} + \mathrm{BC},
	\label{eq:Mbol}
\end{equation}
where the bolometric correction $\mathrm{BC} = 0.128 \log P + 0.022$.

Ultimately, two new values of $Q$ based on $f_0$ were derived: $0.0370 \pm 0.0049 \ \mathrm{d}$ and $0.0271 \pm 0.0011 \ \mathrm{d}$, whose uncertainties are estimated based on the parameters from the TESS Input Catalog and Gaia DR3. Both values fall within the range indicative of the fundamental mode for such stars ($\geq \ 0.027 \ \mathrm{d}$) \citep{Poro2024}.

However, the $Q$ values derived here cannot exclude the possibility that the dominant mode could be the first overtone mode, even more so that the period ratio of F14 ($f_2$) to F1 ($f_0$) is indicative of the period ratio of the second to first radial overtone in HADS. It needs to be addressed in detail with the help of the asteroseismology models in a later part of this work.

In Table \ref{tab:pulsation_constant}, we have also compiled results from historical literature.

\begin{deluxetable*}{l|cccc|cccc}[hbtp]
\label{tab:pulsation_constant}
%\tablewidth{\textwidth}
   \centering
	\tablecaption{Pulsation constant from different sources.}
	\tabletypesize{\footnotesize}
    \tablehead{\colhead{} \vline& \multicolumn{4}{c}{Observed quantities} \vline & \multicolumn{4}{c}{Theoretically derived parameters} \\ 
    \hline
    \colhead{Source}\vline & \colhead{$\Teff \ (\mathrm{K})$} & \colhead{$\logg \  (\mathrm{cm/s}^2$)} & \colhead{Distance (pc)} & \colhead{$V$ (mag)} \vline & \colhead{$M_\mathrm{bol}$ (mag)} & \colhead{$M_{V}$ (mag)} & \colhead{BC} & \colhead{$Q$ (d)}}
    \startdata
    TESS  & 7351.0  & 4.2438  & 422.9060  & 10.99 & 2.4993  & 2.6328  & $-0.1334$  & $0.0370 \pm 0.0049$  \\
    Gaia DR3 & 7153.2   & 3.9766  & 411.8049  & 11.04\tablenotemark{a} & 2.6071  & 2.7405  & $-0.1334$  & $0.0271 \pm 0.0011$  \\				
   \hline
   \citet{McNamara1978} & 7930 & 4.13 & --- & --- & 2.4 & --- & --- & 0.0330\tablenotemark{b}\\
   \citet{McNamara1996} & 7740  & 4.04 & --- & --- & 2.43 & 2.47 & --- & 0.0313\tablenotemark{b}\\
   \citet{pena1999} & 7500 & 3.91 & --- & 10.7459 & --- & --- & --- & 0.0200\tablenotemark{b}\\
\enddata
\tablecomments{		
\tablenotetext{a}{Here, the $V$ band magnitude for CY Aqr comes from the AAVSO Photometric All-Sky Survey (APASS) - Data Release, which was used to fit the Gaia model \citep{Henden2015}.}
\tablenotetext{b}{All these values come from the References directly.}
}
\end{deluxetable*}

\subsection{$O-C$ Analysis}
Given that the amplitude of $f_0$ exceeds that of the other four independent frequencies by approximately two orders of magnitude and the light curves of CY Aqr are dominated by $f_0$, we employed the classical $O-C$ analysis to glean detailed insights into the primary star's period variations through the times of maximum light (TML) of the light curves.

We sourced the TML from four distinct origins:  \citet{fang2016},\citet{Wiedemair2018, Wiedemair2020}, TESS, and the American Association of Variable Star Observers (AAVSO). Following the elimination of outlier data points, all datasets were converted to Barycentric Julian Dates (BJD) utilizing the Astropy Time package\footnote{\url{https://docs.astropy.org/en/stable/time/index.html}} and an online conversion tool\footnote{\url{https://astroutils.astronomy.osu.edu/time/}, \citet{Eastman2010}}.

In total, we collected 2228 TML records, covering a period of approximately 90 years from 1934 to 2023 (refer to Table \ref{tab:TML} in Appendix \ref{app:05}), which includes 1367 TML newly acquired in addition to those reported by \citet{fang2016}.
This extensive dataset offers an unparalleled opportunity to investigate the period variation of the primary star and potentially the properties of its companions.

Specifically, 864 TML were directly extracted from \citet{fang2016}, with three TML instances (HJD: 2439374.6443, 2438660.0051, and 2443815.3636) excluded due to significant deviations. An additional 41 TML were sourced from \citet{Wiedemair2018, Wiedemair2020}. Each of the 1142 TML from the three TESS Sectors was obtained by a fourth-order polynomial fitting to the light curves, with uncertainties determined through Monte Carlo simulation. Furthermore, 184 new TML were derived from AAVSO data spanning 2003 to 2023, using the same methodology as for the TESS data.

For all TML data lacking inherent uncertainties, we assigned uncertainties of $20\ \sigma_\mathrm{TESS}$ for visual photometry, $5\ \sigma_\mathrm{TESS}$ for photographic observations, and $2\ \sigma_\mathrm{TESS}$ for photoelectric and CCD data (where $\sigma_\mathrm{TESS} = 0.00005$ is the mean uncertainty of the TML from TESS), which were then incorporated into the global fitting procedure.

The $O-C$ diagram presented in \citet{fang2016} suggests that CY Aqr is most likely a triple system, eccentrically orbited by two low-mass companions. We fitted the TML using a quadratic function plus two sine terms, which accounts for the LTTE from two companions in independent elliptical orbits. The resulting formula for the calculated TML is:
\begin{equation}
  \label{eq:oc-fitted}
  \begin{split}
  C =\quad & \mathrm{BJD}_0 + P_0 \times E + \frac{1}{2} \beta E^2 + \\
  & \sum_{k=\mathrm{A,B}} A_{k}\left[\sqrt{1-e_{k}^2}\sin{\phi_{k}} \cos{\omega_{k}}+\cos{\phi_{k}}\sin{\omega_{k}}\right],
  \end{split}
\end{equation}
where $\phi_{k}$ ($k \equiv \mathrm{A, B}$) is the solution of Kepler's equation:
\begin{equation}
  \label{eq:phi}
  \phi_{k} - e_{k} \sin \phi_{k} = \frac{2 \pi}{P_{\mathrm{orb,k}}}(P_0 \times E - T_{0,k}).
\end{equation}

In these equations, $\text{BJD}_0$ denotes the initial epoch based on Barycentric Julian Day, $P_0$ is the pulsation period of the primary star, $\beta$ represents the linear variation of the pulsation period of the primary star, $A_{k}=a_{k}\sin{i}_{k}/c$ (with $c$ being the speed of light in vacuum) is the projected semi-major axis of the companion's orbit, $e_{k}$ is the eccentricity of the companion's orbit, $\phi_{k}$ is the eccentric anomaly of the companion's orbit, $\omega_{k}$ (the argument of periastron) is the angle from the ascending node to periastron in the companion's orbital plane, $P_{\text{orb,k}}$ is the orbital period of the companions, and $T_{0,k}$ is the time of passage through the periastron of the companions. For more details on the light-time orbit equation, refer to \citet{Xue2020}.

\begin{figure*}[!htbp]
	\centering
	\includegraphics[width=0.8\textwidth]{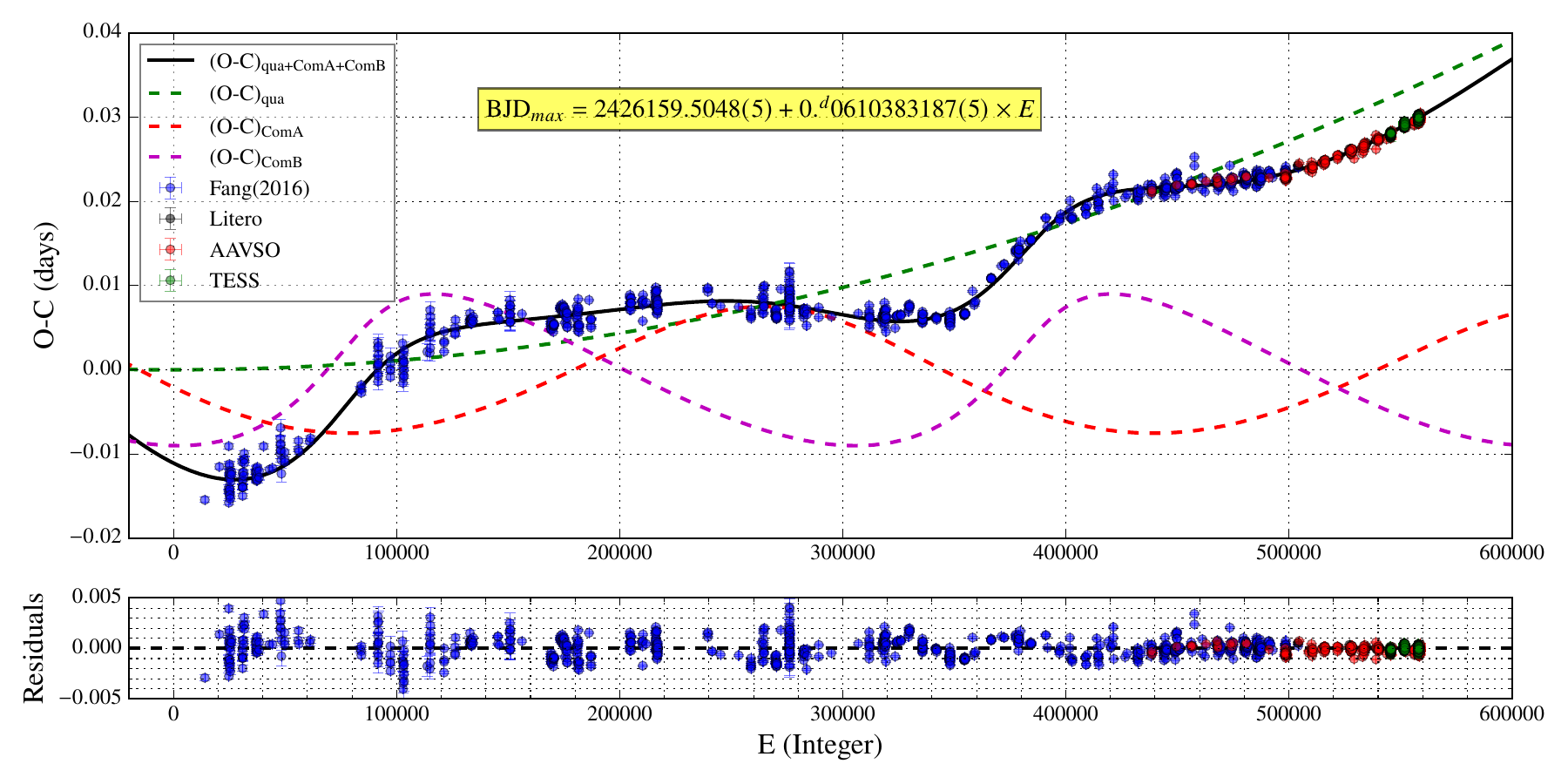}
\caption{$O-C$ values (excluding the linear part BJD$_{\text{max}}$) and the corresponding residuals. In the upper panel, the black line signifies the best-fit result of a quadratic plus two light-time orbital terms, with the green dashed line indicating the quadratic component, { the red dashed line indicating the light-time orbital term from Companion A, and the magenta dashed line indicating the light-time orbital term from Companion B}. The lower panel displays the residuals of the best-fit result. Data from \citet{fang2016} are represented by blue points; historical literature data \citep{Wiedemair2018,Wiedemair2020} by grey points; AAVSO data by red points; and TESS data by green points.
} 
\label{fig:OC_CYAqr}
\end{figure*}

The Markov Chain Monte Carlo (MCMC) algorithm was employed to ascertain the posterior probability distribution of the parameters in Eqs. (\ref{eq:oc-fitted}) and (\ref{eq:phi}).\footnote{The {\sc python} module {\tt emcee} \citep{Foreman2013} is employed to perform the MCMC sampling. Some examples can be found in \citet{Niu201801,Niu201802,Niu2019,Niu2022CRs} and references therein.} The samples of the parameters were taken from their posterior probability distribution function (PDF) once the Markov Chains reached equilibrium. The mean values and standard deviations of the parameters are tabulated in Table~\ref{tab:pul_orb_para}, and the best-fit result (yielding $\chi^{2}/\text{d.o.f.} = 47.05$) for the $O-C$ values (excluding the linear part) along with the corresponding residuals are depicted in Figure~\ref{fig:OC_CYAqr}.

\begin{deluxetable}{ccc}
	\label{tab:pul_orb_para}
	\tablecaption{Pulsating and orbital parameters of CY Aqr.}
	\tabletypesize{\footnotesize}
	\tablehead{
		\colhead{Parameter} & \multicolumn{2}{c}{Value} } 
	\startdata
	$\BJD_0$  & \multicolumn{2}{c}{$2426159.50485 \pm 0.00003$} \\
	$P_0 \ (\mathrm{d})$    & \multicolumn{2}{c}{$0.06103831875 \pm 0.00000000006$} \\
	$\beta \ (\mathrm{d \ cycle}^{-1})$  & \multicolumn{2}{c}{$(2.175 \pm 0.002) \times 10^{-13}$ } \\
	$\dPdt \ (\mathrm{yr^{-1}})$ & \multicolumn{2}{c}{$(2.132 \pm 0.002) \times 10^{-8}$} \\
	\hline
	Parameter & Companion A & Companion B \\
	\hline
	$A \ (\mathrm{d})$ & $0.0076 \pm 0.0001$ & $0.0098 \pm 0.0001$ \\
	$e$      & $0.139 \pm 0.002$ & $0.431 \pm 0.004$ \\
	$\Porb\ (\mathrm{d})$ & $21990.6 \pm 62.9$ & $18547.0 \pm 25.5$ \\
	$T_0\ (\mathrm{BJD})$    & $2421432.6 \pm 164.8$ & $2431156.4 \pm 25.8$ \\
	$\omega\ (^{\circ})$ & $-249.0 \pm 1.6$ & $24.5 \pm 0.2$ \\
	\hline
	$a \sin i\ (\mathrm{AU})$   & $1.32 \pm 0.02$ & $1.70 \pm 0.02$ \\
	$f(m) \ (\Msun)$ & $0.00063 \pm 0.00003$ & $0.00190 \pm 0.00006$\\
	\enddata
\end{deluxetable}

\section{Discussion and Conclusion} 
\label{sec:dis_con}

\subsection{About the Primary Star}
The asteroseismological models of the primary star of CY Aqr were constructed using the Modules for Experiments in Stellar Astrophysics (MESA; \citep{Paxton2011, Paxton2013, Paxton2015, Paxton2018, Paxton2019, Jermyn2023}) and the stellar oscillation code GYRE \citep{Townsend2013}. Two metallicity values, $\mathrm{[Fe/H]} = -0.7$ \citep{Rodriguez1990} and $\mathrm{[Fe/H]} = -1.96$ \citep{Gaia2021}, corresponding to $Z$ values of $0.003$ and $0.00015$ \citep{Niu2022}, respectively, were incorporated into our calculation.

\begin{figure*}[!htbp]
	\centering
	\includegraphics[width=0.49\textwidth]{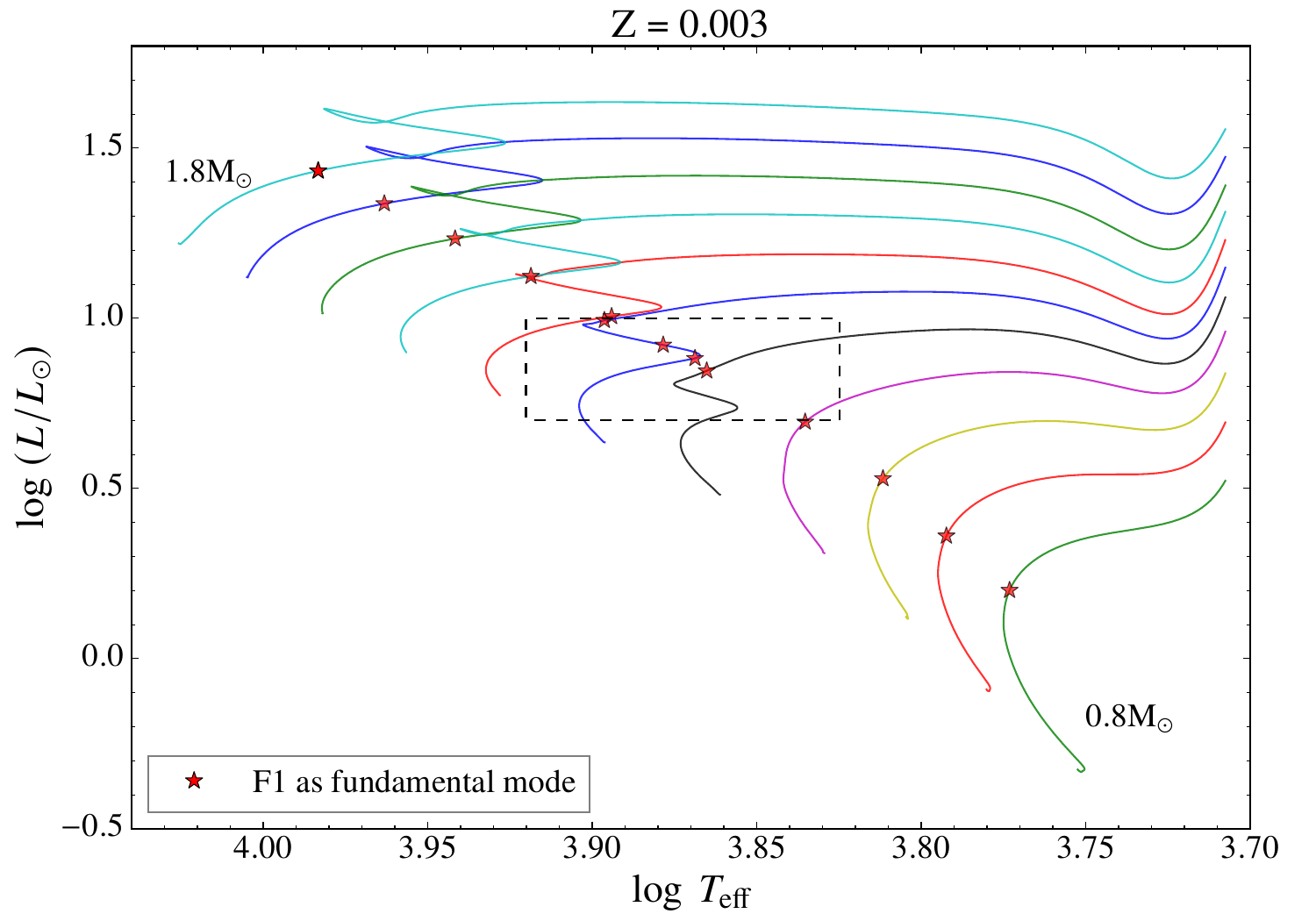}
	\includegraphics[width=0.49\textwidth]{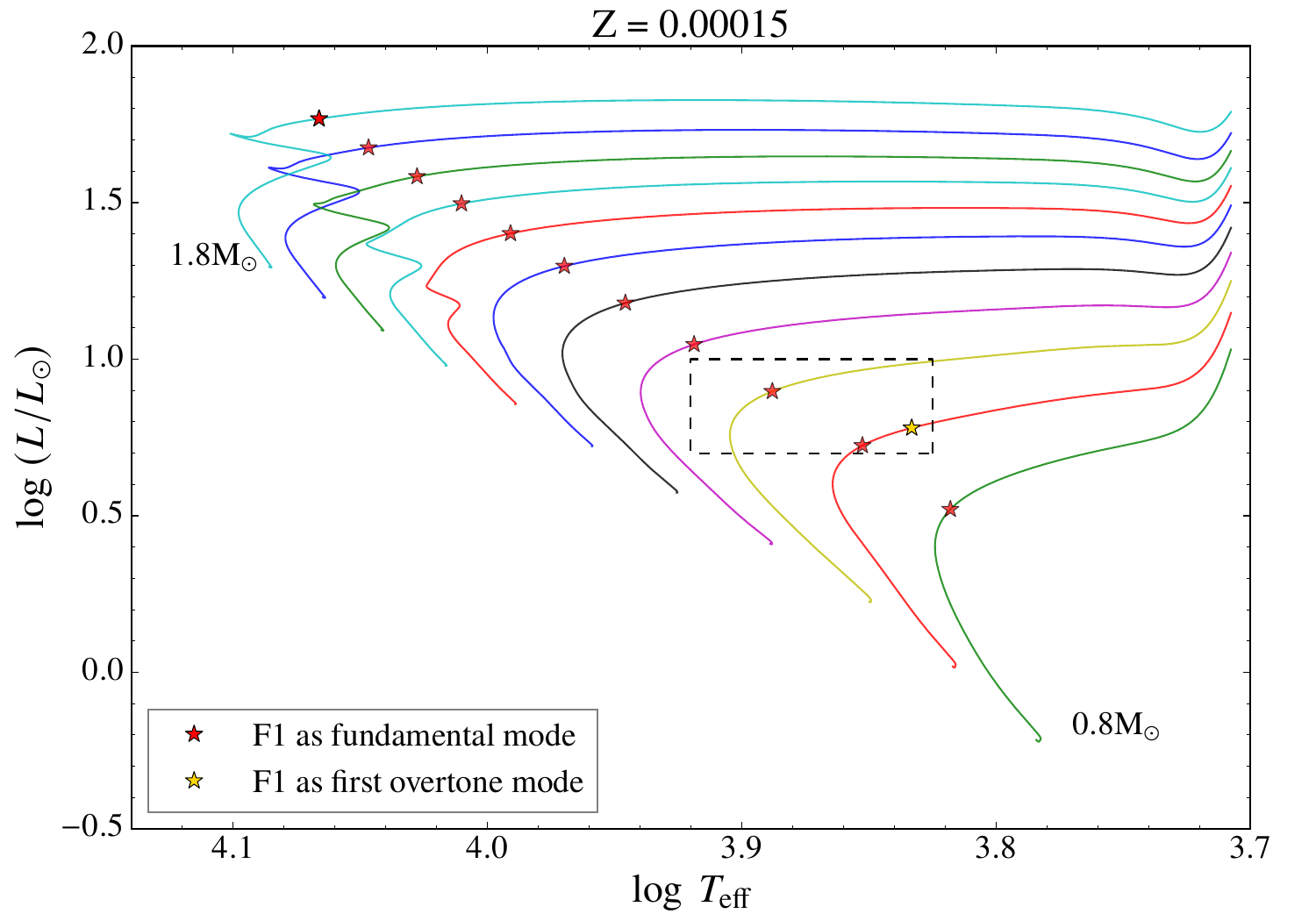}
	\caption{Best-fit seismic models superimposed on evolutionary tracks. The left and right panels display results for $Z=0.003$ and $Z=0.00015$, respectively. The best-fit models are indicated by pentagrams, and the dashed line boxes outline the observed luminosity and effective temperature ranges.}
	\label{fig:hr}
\end{figure*}

The luminosity and effective temperature ranges were determined following the methodology outlined in \citet{Xue2020}, with relevant parameters sourced from \citet{Schlafly2011,fang2016,pena1999} and Gaia DR3. The outcomes are delineated within the dashed line box in Figure \ref{fig:hr}, with $\log L/L_{\odot} \in [0.70,1.00]$ and $\Teff \in [6680,8320]\ \mathrm{K}$.

\begin{figure*}[!htbp]
	\centering
	\includegraphics[width=0.49\textwidth]{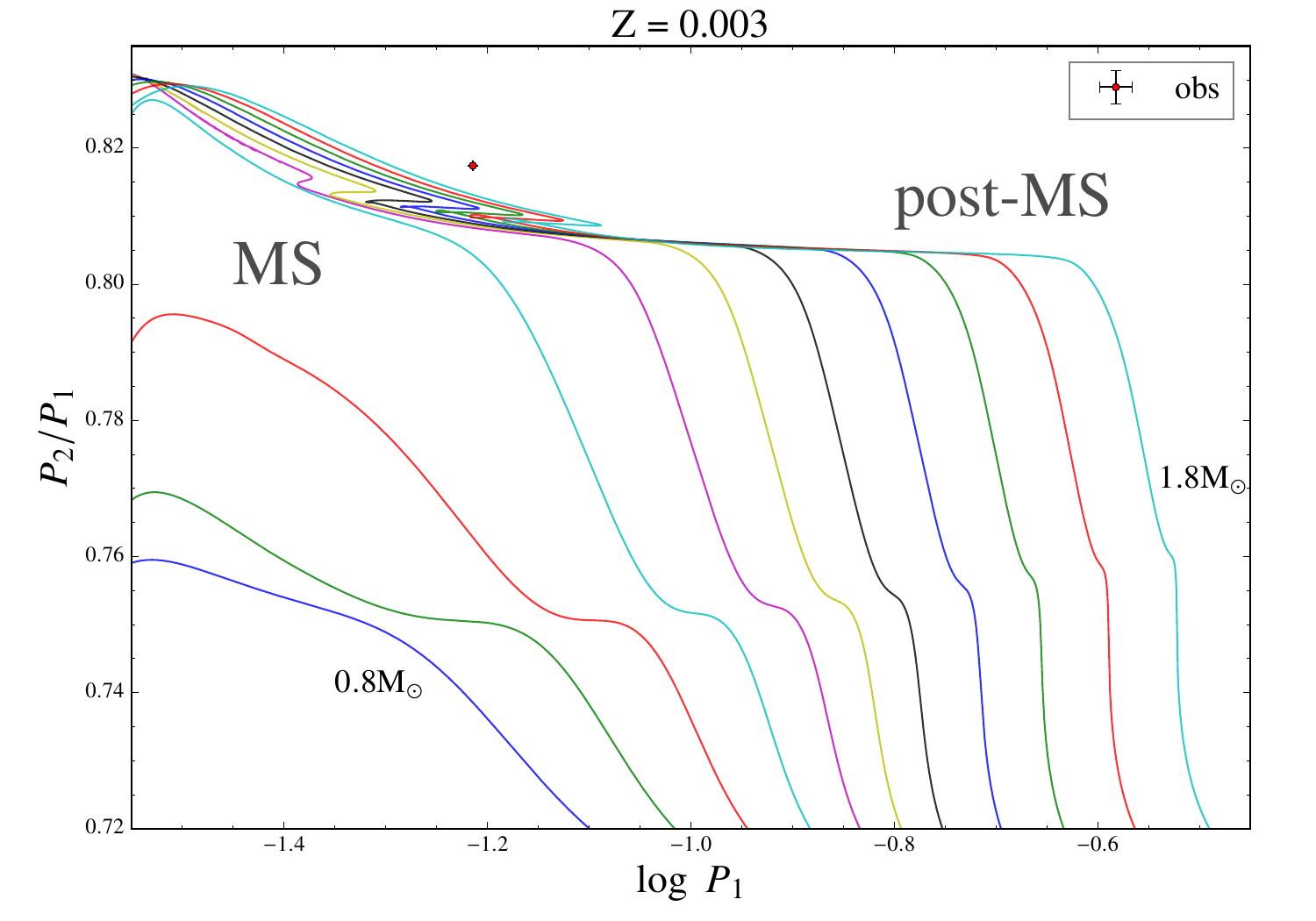}
	\includegraphics[width=0.49\textwidth]{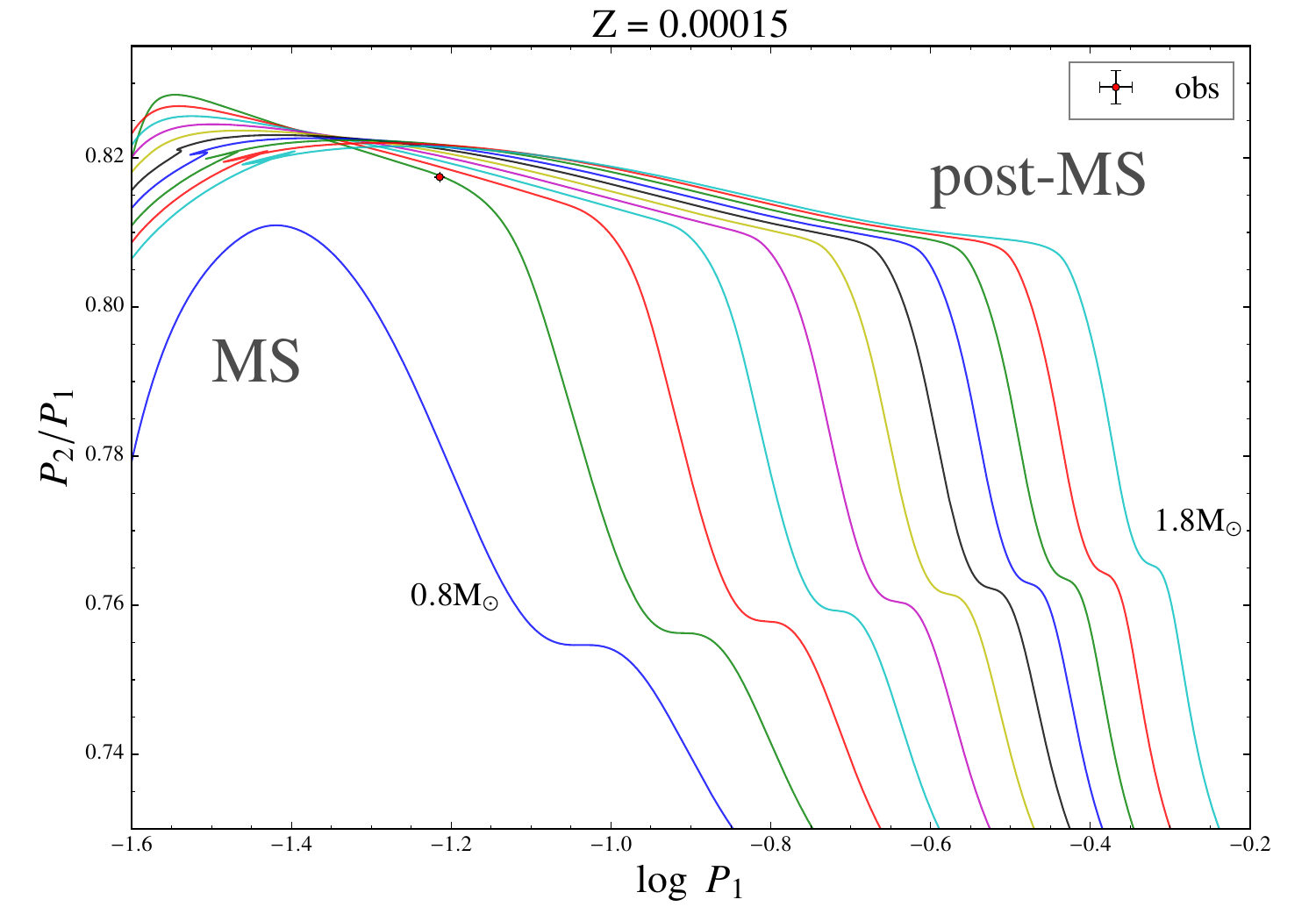}
	\caption{Petersen diagram with the observed value and all evolutionary tracks for the first and second overtone modes. The left and right panels display results for $Z=0.003$ and $Z=0.00015$, respectively. The observed value of first and second overtone (F1 and F14) are indicated by red dot with error bars. $P_1$ and $P_2$ are the periods of the first and second overtone mode (corresponding to F1 and F14), respectively.}
	\label{fig:petersen}
\end{figure*}

As that has been discussed in Sec. \ref{subsec:ModeIden}, we cannot exclude the possibility that the dominant pulsation mode could be the first overtone mode based on $Q$ value alone. Here, we present the Petersen diagram in Figure \ref{fig:petersen} with the help of the theoretical evolutionary tracks, in which we consider F1 ($f_0$) and F14 ($f_2$) as the first and second overtone modes. In Figure \ref{fig:petersen}, the theoretical evolutionary tracks pass through the observed value (the red dot) within uncertainties just when $Z=0.00015$ and $M = 0.9 \Msun$, in which case F1 and F14 can be considered as the first and second overtone modes.

If we consider F1 as the fundamental mode of the primary star, we can determine the mass of it to be within the range of $0.9-1.0\ \Msun$ and $1.2-1.3\ \Msun$ for varying metallicities, by integrating observed constraints from the fundamental frequency, luminosity, and effective temperature.
Figure \ref{fig:hr} illustrates the best-fit seismic models alongside evolutionary tracks for different metallicities. One should note that there exist three different possible evolutionary stages when $Z=0.003$ and $M = 1.3 \Msun$, which correspond to the terminal-age main sequence, the overall contraction and the post main sequence evolutionary stages. 
All the parameters of the best-fit models (within the derived luminosity and effective temperature ranges) are listed in Table \ref{tab:astero_results}.

\begin{deluxetable*}{c|cccccc|c}[!htbp]
	\label{tab:astero_results}
%%	\centering
	\tablecaption{Parameters of the best-fit models.}
%	\tabletypesize{\footnotesize}
	\tablehead{\colhead{} \vline& \multicolumn{6}{c}{F1 as fundamental mode} \vline& \colhead{F1 as first overtone mode} \\ 
	\hline
	\colhead{Mass ($\Msun$)} \vline & \colhead{0.9}  & \colhead{1.0} & \colhead{1.2} & \colhead{1.3\tablenotemark{a}} & \colhead{1.3\tablenotemark{b}} & \colhead{1.3\tablenotemark{c}} \vline &\colhead{0.9} \\ 
	\colhead{$Z$} \vline & \colhead{0.00015}  & \colhead{0.00015} & \colhead{0.003} & \colhead{0.003} & \colhead{0.003} & \colhead{0.003} \vline& \colhead{0.00015}}
    \startdata
%%		$Z$ &0.00015  & 0.00015  & 0.003  & 0.003  \\
		$\Teff$ (K) &7121 & 7726  & 7331  & 7392 & 7558 & 7876 & 6810\\
		$\log L/L_\odot$& 0.72  & 0.90  & 0.85  & 0.88 & 0.92 & 0.99 & 0.78\\
		$f_0 \ (\cd)$& 16.3830  & 16.3831 &16.3831 & 16.3832 & 16.3831 & 16.3832 & 16.3832\tablenotemark{d}\\
		{$\dPdt$} & 1.215 & 1.944 & 0.687 & 0.595 & -2.774 & 10.600 & 2.650\tablenotemark{e}\\
		($\times 10^{-9}yr^{-1}$) & --- & --- & --- & --- & --- & --- & --- \\
\enddata
\tablecomments{		
\tablenotetext{a}{Corresponding to the terminal-age main sequence evolutionary stage.}
\tablenotetext{b}{Corresponding to the overall contraction evolutionary stage.}
\tablenotetext{c}{Corresponding to the post main sequence evolutionary stage.}
\tablenotetext{d}{Here, it is the frequency of the first overtone mode.}
\tablenotetext{e}{Here, it is the period variation rate of the first overtone mode.}
}
\end{deluxetable*}

The linear period variation rates of these fitted seismic models are consistently smaller than the observed rate of  $(2.132 \pm 0.002) \times 10^{-8} \yr^{-1}$ as presented in Table \ref{tab:pul_orb_para}, suggesting that the primary star of CY Aqr may be undergoing mass loss. Employing the method proposed by \citet{Xue2020}, we derived the mass loss rate from the best-fit models detailed in Table \ref{tab:astero_results}, yielding a range from $3.36 \times 10^{-8} \Msun \yr^{-1}$ to $6.26 \times 10^{-8} \Msun \yr^{-1}$.

Based on the mass loss rate in main-sequence B single stars \citep{Massloss2014}, we can obtain an upper limit of it in (post) main sequence A-F single stars as $\sim 10^{-12}-10^{-11} \Msun \yr^{-1}$, which can be ignored comparing to the primary star of CY Aqr.
{ Because of the lack of direct observational evidence, here we propose three possible origins of the substantial mass loss of the primary star: (i) the pulsation induced mass loss. Although this mechanism is believed to occur in some Cepheids \citep{Wilson1984}, it is not yet well understood and needs more observational evidences in SX Phe stars; (ii) the mass transfer between the primary star and companions in the evolutionary history. The mass transfer will distrub the normal evolution of the primary star as a single star, which would lead to an inaccurate estimation of the evolution induced period variaiton rate, and then the mass loss rate; (iii) the complexity of three-body problem. Strictly speaking, the LTTE model we used in this work (a simple superposition of two two-body systems) is just an approximation of the real three-body system, which might be a good approximation in short time scale, but incorrect in long time scale. It would also lead to an inaccurate estimation of $\dPdt$ and then the mass loss rate.}

%%Consequently, this rate is likely closely associated with its two companion stars.

\subsection{About the Two Companions}
With the expansion of the TML and the refinement of our numerical fitting frameworks, we have, for the first time, determined the time of periastron passage $T_0$ and the longitude of the periastron $\omega$ for both companions.
Moreover, we derived the mass functions for Companion A and B as $f(m_\mathrm{A}) = 0.00063 \pm 0.00003 \ \Msun$ and $f(m_\mathrm{B}) = 0.00190 \pm 0.00006  \ \Msun$, respectively (refer to Table \ref{tab:pul_orb_para}), which are at least an order of magnitude larger than those obtained in \citet{fang2016}. 
{ The differences could come from the employment of the quadratic term in Eq. \ref{eq:oc-fitted} (which is significantly non-zero), a more complete LTTE model, and the newly extended data in this work. Moreover, we suggest that a global fitting to all the free parameters simultaneously in such problem is needed, which can help us to obtain a global optimum solution rather than local optimum solutions in most cases.}

Assuming no interactions between the two companions and based on the mass function expression,
\begin{equation} 
\label{eq:mass_function}
	f(m_k) = \frac{4\pi^2 (a_k \sin i_k)^3}{G P_{\mathrm{orb},k}^2}  = \frac{(m_k \sin i_k)^3}{(M + m_k)^2},
\end{equation}
we can independently determine the relationship between $i_k$ and $m_k$ for Companion A and B (for a given $M$).
Setting $M = 0.9 \Msun, 1.0 \Msun, 1.2 \Msun, 1.3 \Msun$ and $i_k \in [10^{\circ}, 90^{\circ}]$, we find the mass range for the companions to be $m_{\mathrm{A}} \in [0.085, 0.818] \Msun$ and $m_{\mathrm{B}} \in [0.126, 1.382] \Msun$.

Considering the weak hydrogen and metallic lines near minimum light \citep{Struve1949} and the mean reddening-free color indices $(b - y)_0 \in (0.061-0.211)$ (corresponding to spectral types A to F) \citep{McNamara1996}, CY Aqr's companions cannot be bright red stars (which will give a obvious hydrogen or metallic lines near minimum light and significant larger value of $(b - y)_0$ \citep{Fu2008}), but low-luminosity degenerate stars, such as white dwarfs, consistent with the derived mass range.

\subsection{About the Triple System}
However, noting that $P_{\mathrm{orb,A}} > P_{\mathrm{orb,B}}$ but $a_{\mathrm{A}} \sin i_{\mathrm{A}} < a_{\mathrm{B}} \sin i_{\mathrm{B}}$, it suggests that the orbital inclination between Companion A and B is significant and provides an additional constraint on $(i_{\mathrm{A}}, m_{\mathrm{A}})$.

Specifically, the orbital period and the semi-major axis of the orbit for Companion A and B can be expressed as
\begin{equation} 
\label{eq:P_a}
	P_{\mathrm{orb},k}^2 = \frac{4\pi^2}{G(M + m_k)} a_k^3,
\end{equation}
where $G$ is the gravitational constant, $M$ is the mass of the primary star, and $m_k$ is the mass of the companions.

Based on Eq. (\ref{eq:P_a}), we derive the orbit inclination relation between the two companions as
\begin{equation} 
\label{eq:i_AB}
\begin{split}
	\frac{\sin i_{\mathrm{A}}}{\sin i_{\mathrm{B}}} &= \frac{a_\mathrm{A} \sin i_{\mathrm{A}}}{a_\mathrm{B} \sin i_{\mathrm{B}}} \left( \frac{P_{\mathrm{orb,B}}}{P_{\mathrm{orb,A}}} \right)^{\frac{2}{3}} \cdot \left(\frac{M+m_{\mathrm{B}}}{M+m_{\mathrm{A}}} \right)^{\frac{1}{3}} \\ 
	&\approx 0.69 \cdot \left(\frac{M+m_{\mathrm{B}}}{M+m_{\mathrm{A}}} \right)^{\frac{1}{3}},
\end{split}
\end{equation}
which establishes a relationship between $(i_{\mathrm{A}}, m_{\mathrm{A}})$ and $(i_{\mathrm{B}}, m_{\mathrm{B}})$ for a given $M$.

In this scenario, for a specified $M$, assigning a particular value to $i_{\mathrm{B}}$ yields the corresponding $m_{\mathrm{B}}$ from Eq. (\ref{eq:mass_function}) (for Companion B). By combining Eqs. (\ref{eq:mass_function}) (for Companion A) and (\ref{eq:i_AB}), we can also determine the corresponding values of $(i_{\mathrm{A}}, m_{\mathrm{A}})$.
Thus, the arrays $(i_{\mathrm{A}}, m_{\mathrm{A}})$ and $(i_{\mathrm{B}}, m_{\mathrm{B}})$ exhibit a one-to-one correspondence, significantly constraining the mass and orbital inclination of Companion A.

\begin{figure*}
	\centering
	\includegraphics[width=0.49\textwidth]{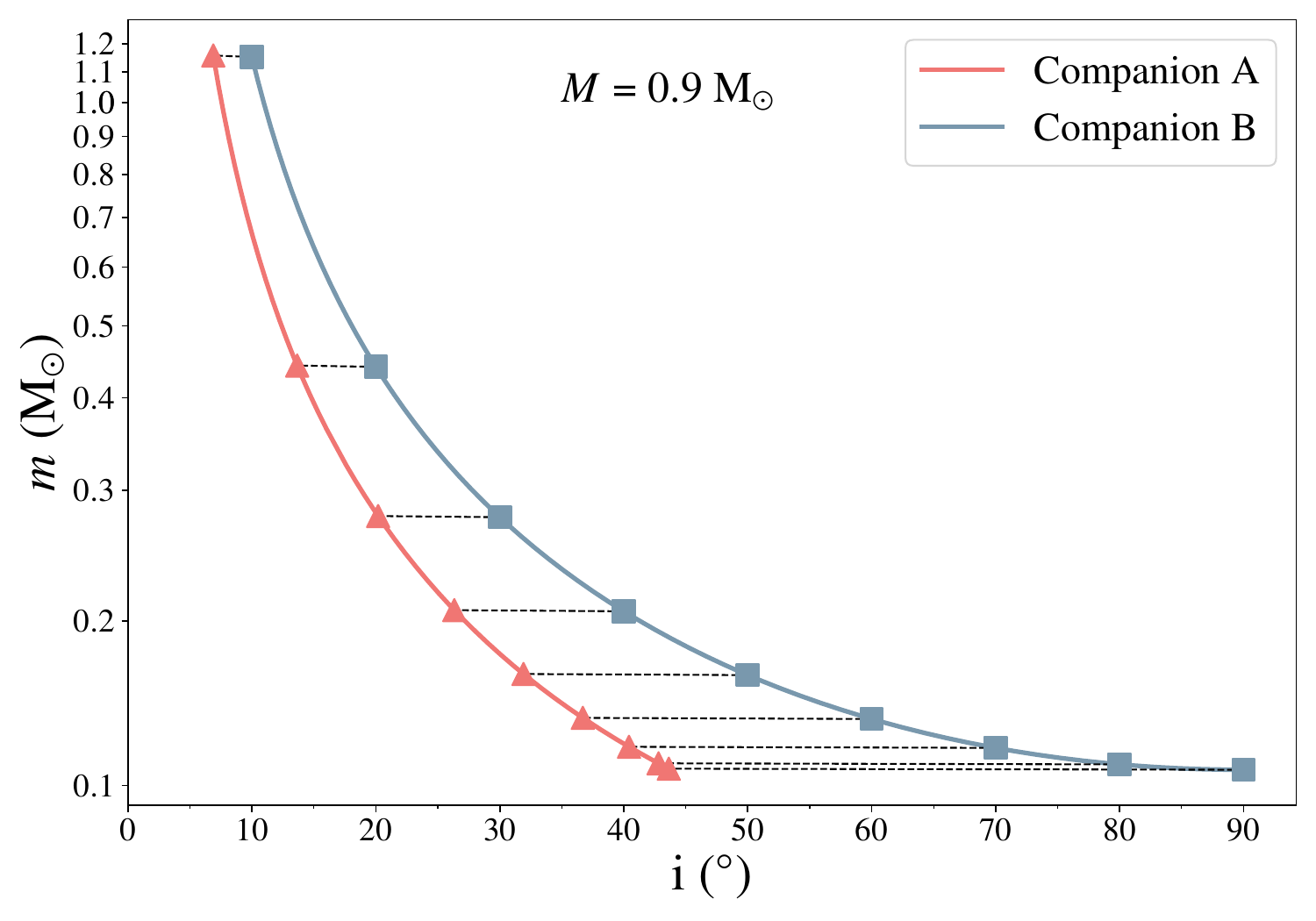}
	\includegraphics[width=0.49\textwidth]{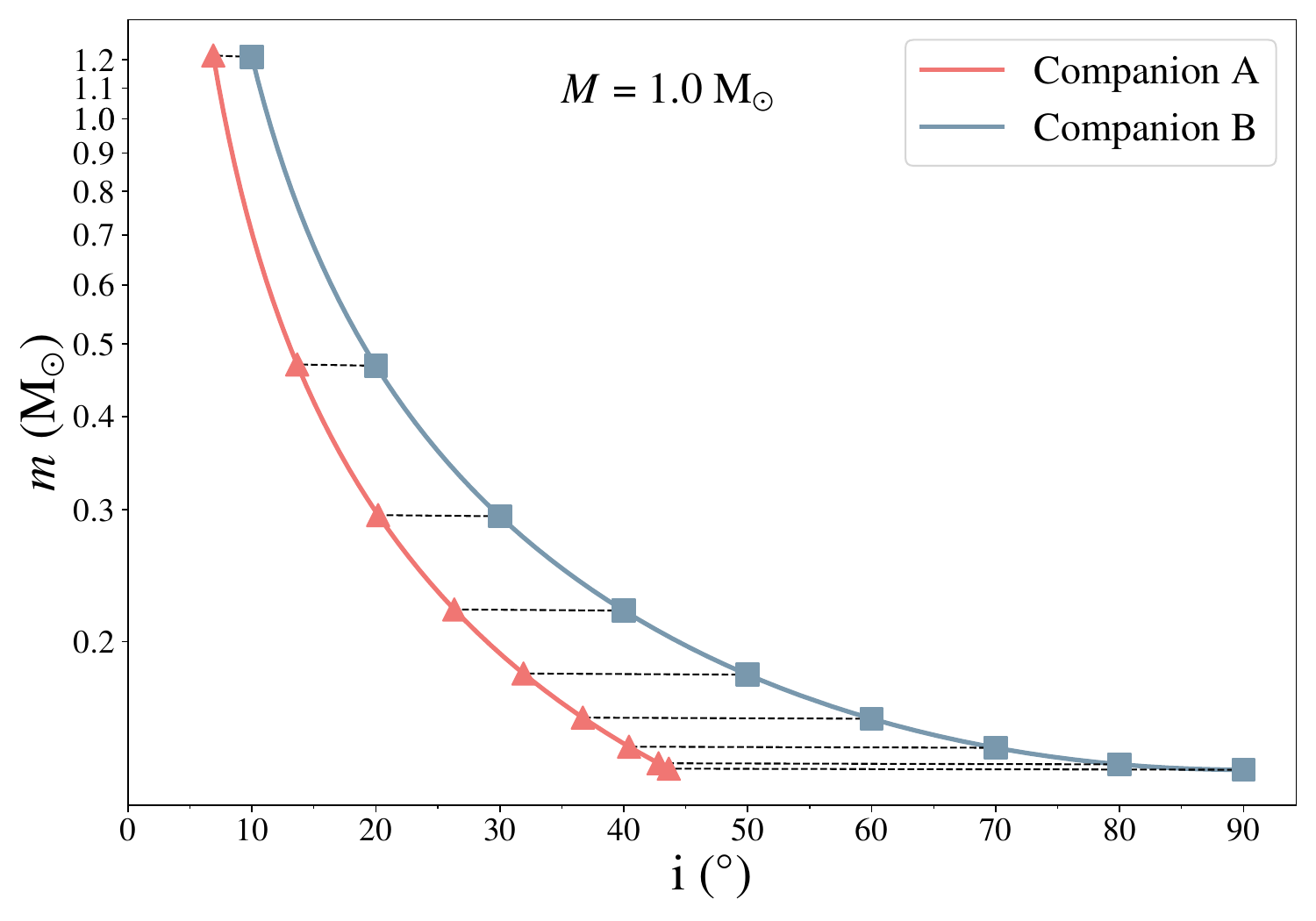}
	\includegraphics[width=0.49\textwidth]{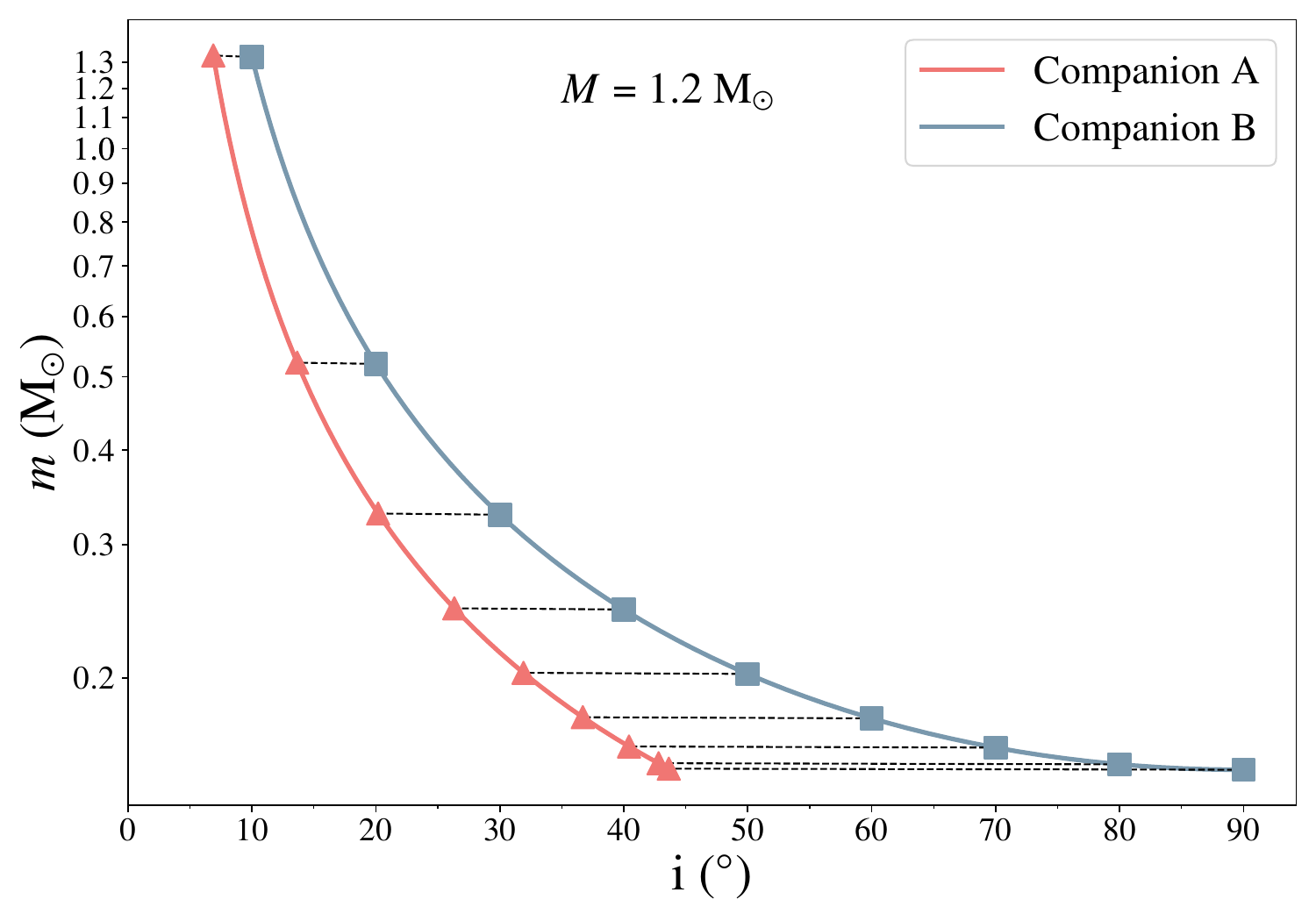}
	\includegraphics[width=0.49\textwidth]{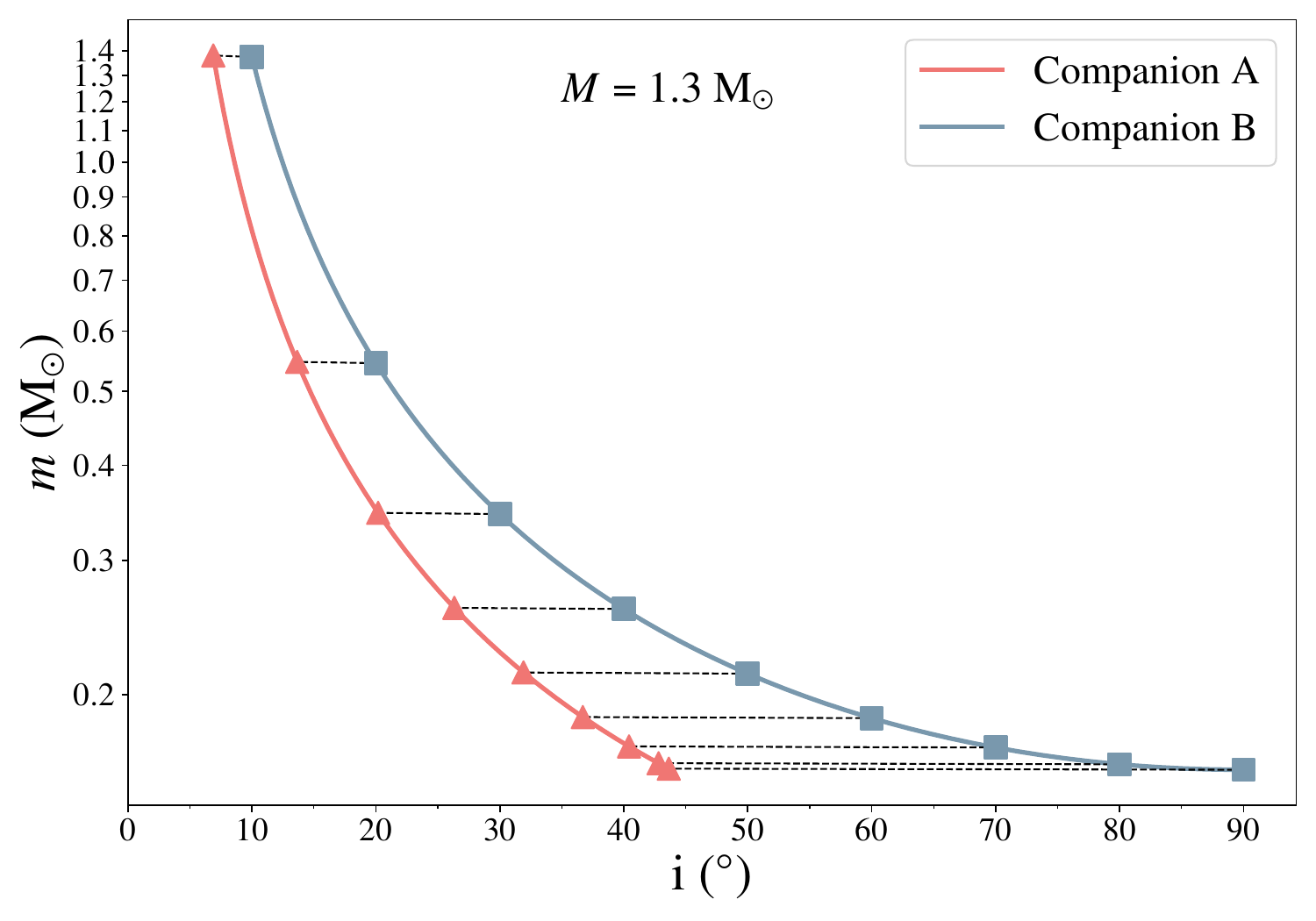}
	\caption{Relationship between the mass and the orbit inclination for Companions A and B when the mass of the primary star is $M = 0.9 \Msun, 1.0 \Msun, 1.2 \Msun, 1.3 \Msun$. In each subfigure, the red line and the dark blue line represent the relationships of $(i_{\mathrm{A}}, m_{\mathrm{A}})$ and $(i_{\mathrm{B}}, m_{\mathrm{B}})$, respectively. The dark blue squares represent the anchor points when $i_{\mathrm{B}} = 10^{\circ}, 20^{\circ}, 30^{\circ}, 40^{\circ}, 50^{\circ}, 60^{\circ}, 70^{\circ}, 80^{\circ}, 90^{\circ}$, and the red triangles represent the corresponding points for Companion A, connected by dashed lines.}
	\label{fig:mass_AB}
\end{figure*}

In Table \ref{tab:mass_AB}, we present the one-to-one correspondence values of $(i_{\mathrm{A}}, m_{\mathrm{A}})$ and $(i_{\mathrm{B}}, m_{\mathrm{B}})$ for $i_{\mathrm{B}}$ ranging from $10^{\circ}$ to $90^{\circ}$ and primary star masses of $M = 0.9\ \Msun, 1.0\ \Msun, 1.2\ \Msun, 1.3\ \Msun$.
Based on the mass functions (Eq. (\ref{eq:mass_function})), the relationships between $(i_{\mathrm{A}}, m_{\mathrm{A}})$ and $(i_{\mathrm{B}}, m_{\mathrm{B}})$ are depicted in Figure \ref{fig:mass_AB}.
It is evident that the orbital inclination $i_{\mathrm{A}} \le 43.64^{\circ}$, imposing a stringent constraint on the configuration of this triple system.

\centerwidetable
\begin{deluxetable*}{cc|cc|cc|cc|cc|cc|cc|cc}
    \label{tab:mass_AB}
    \tablecaption{Orbital inclinations and companion masses of Companion A and B for different primary star masses.}
    \tablewidth{0pt}
%    \tabletypesize{\small}
    \tablehead{
        \multicolumn{4}{c}{$M = 0.9 \Msun$} \vline  &  \multicolumn{4}{c}{$M = 1.0 \Msun$} \vline &  \multicolumn{4}{c}{$M = 1.2 \Msun$}  \vline &  \multicolumn{4}{c}{$M = 1.3 \Msun$} \\ \hline \colhead{$i_\mathrm{A}$} & \colhead{$m_\mathrm{A}$} \vline& \colhead{$i_\mathrm{B}$} & \colhead{$m_\mathrm{B}$} \vline&  \colhead{$i_\mathrm{A}$} & \colhead{$m_\mathrm{A}$} \vline& \colhead{$i_\mathrm{B}$} & \colhead{$m_\mathrm{B}$} \vline&  \colhead{$i_\mathrm{A}$} & \colhead{$m_\mathrm{A}$} \vline& \colhead{$i_\mathrm{B}$} & \colhead{$m_\mathrm{B}$} \vline&
        \colhead{$i_\mathrm{A}$} & \colhead{$m_\mathrm{A}$} \vline& \colhead{$i_\mathrm{B}$} & \colhead{$m_\mathrm{B}$} } 
    \startdata
    6.89  & 1.165  & 10    & 1.158  & 6.89  & 1.224  & 10    & 1.217  & 6.89  & 1.336  & 10    & 1.329  & 6.89  & 1.389  & 10    & 1.382  \\
    13.66  & 0.444  & 20    & 0.442  & 13.66  & 0.472  & 20    & 0.470  & 13.66  & 0.525  & 20    & 0.522  & 13.66  & 0.550  & 20    & 0.547  \\
    20.19  & 0.279  & 30    & 0.277  & 20.19  & 0.297  & 30    & 0.295  & 20.19  & 0.332  & 30    & 0.330  & 20.19  & 0.348  & 30    & 0.347  \\
    26.34  & 0.208  & 40    & 0.207  & 26.34  & 0.222  & 40    & 0.221  & 26.34  & 0.249  & 40    & 0.247  & 26.34  & 0.261  & 40    & 0.260  \\
    31.92  & 0.171  & 50    & 0.170  & 31.92  & 0.182  & 50    & 0.181  & 31.92  & 0.204  & 50    & 0.203  & 31.92  & 0.215  & 50    & 0.214  \\
    36.70  & 0.149  & 60    & 0.148  & 36.70  & 0.159  & 60    & 0.158  & 36.70  & 0.179  & 60    & 0.178  & 36.70  & 0.188  & 60    & 0.187  \\
    40.43  & 0.136  & 70    & 0.135  & 40.43  & 0.145  & 70    & 0.145  & 40.43  & 0.163  & 70    & 0.163  & 40.43  & 0.172  & 70    & 0.171  \\
    42.82  & 0.129  & 80    & 0.129  & 42.82  & 0.138  & 80    & 0.138  & 42.82  & 0.155  & 80    & 0.155  & 42.82  & 0.163  & 80    & 0.163  \\
    43.64  & 0.127  & 90    & 0.126  & 43.64  & 0.136  & 90    & 0.135  & 43.64  & 0.153  & 90    & 0.152  & 43.64  & 0.161  & 90    & 0.160  \\
\enddata
\tablecomments{Unit of $i_\mathrm{A}$ and $i_\mathrm{B}$: $^\circ$; unit of $m_\mathrm{A}$ and $m_\mathrm{B}$: $\Msun$.}
\end{deluxetable*}

\begin{figure}[!htbp]
	\centering
	\includegraphics[width=0.49\textwidth]{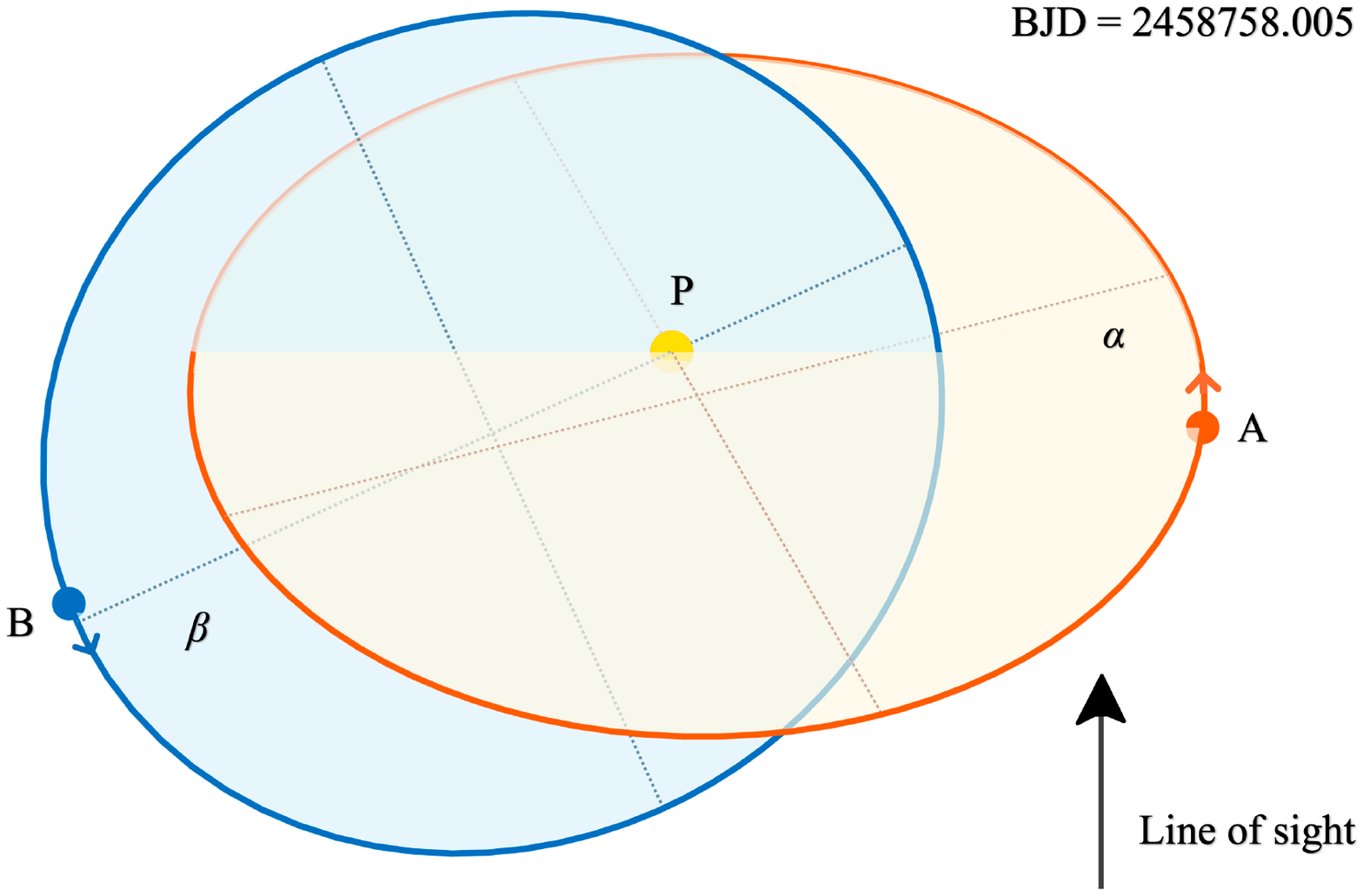}
	\caption{Positions of the three stars and the relationship between the orbits of Companion A and B when $i_{\mathrm{B}} = 90^{\circ}$ and $\mathrm{BJD} = 2458758.005$. P, A, and B denote the primary star, Companion A, and Companion B, respectively. $\alpha$ and $\beta$ denote the orbital planes of Companion A and B.} 
	\label{fig:orbits}
\end{figure}

In a particular scenario where $i_{\mathrm{B}} = 90^{\circ}$ and $\mathrm{BJD} = 2458758.005$, Figure \ref{fig:orbits} illustrates the positions of the three stars and the relationship between the orbits of Companion A and B\footnote{Here, we also assume that the orbital planes of Companion A and B has an identical longitude of ascending node.}, which shows a nested sturcture between the orbitals.

Given that the eccentricity of Companion A ($0.139 \pm 0.002$) is less than that of Companion B ($0.431 \pm 0.004$) and considering the significant inclination between their orbits, we can reasonably infer that the primary star and Companion A were co-evolving, while Companion B was likely captured by the primary star, subsequently intruding into the orbit of Companion A.
Furthermore, considering that the eccentricity of Companion A is the smallest among similar multiple systems with confirmed eccentricities\footnote{The similar multiple systems are collected in Table \ref{tab:multiple_sys}, Appendix \ref{app:05}, see more details therein.}, CY Aqr stands out as an exceedingly rare case.

However, what is even more remarkable is that the mass of Companion A is consistently slightly larger than that of Companion B, with differences of less than 1\% of $m_{\mathrm{A}}$ or $m_{\mathrm{B}}$ (as seen in Table \ref{tab:mass_AB}). When accounting for the uncertainties of the orbital parameters, Companions A and B effectively possess identical masses.
Though this could be dismissed as a mere coincidence, such an explanation seems implausible.

If we believe that the system represents an inevitable and special evolutionary stage in which Companion A and B have exact masses, the evolutionary history and mass redistribution mechanism of such system emerge as interesting enigmas.
Further detailed observations of CY Aqr will be instrumental in unraveling these enigmas in the future.

\section*{Acknowledgments}
H.F.X. acknowledges support from the National Natural Science Foundation of China (NSFC) (No. 12303036). 
All the authors acknowledge the TESS Science team and everyone who has contributed to making the TESS mission possible. We also acknowledge with thanks the variable star observations from the AAVSO International Database contributed by observers worldwide and used in this research.

All the {\it TESS} data used in this paper can be found in MAST:  \dataset[10.17909/3GQN-PZ22]{http://dx.doi.org/10.17909/3gqn-pz22}.

%\object{CY Aquarii}
\facility{AAVSO}
\software{{\tt astropy} \citep{Astropy2013,Astropy2018,Astropy2022}, {\tt Lightkurve} \citep{lightkurve}, {\tt NumPy} \citep{numpy}, {\tt SciPy} \citep{scipy}, {\tt matplotlib} \citep{matplotlib}, {\tt emcee} \citep{Foreman2013}, {\tt Period04} \citep{Lenz2005}, {\tt MESA} \citep{Paxton2011, Paxton2013, Paxton2015, Paxton2018, Paxton2019, Jermyn2023}, {\tt GYRE} \citep{Townsend2013}}

%%\bibliography{ref}

\setcounter{figure}{0}
\setcounter{table}{0}
\renewcommand{\thefigure}{A\arabic{figure}}
\renewcommand{\thetable}{A\arabic{table}}

\appendix

\section{Related works}
\label{app:05}
We have collected the similar systems studied by $O-C$ analysis in Table \ref{tab:multiple_sys}. More details can be referred to the original references.

\begin{table*}[!htp] 
\begin{center}
\caption{Similar systems studied by $O - C$ analysis.\label{tab:multiple_sys}}
%	\tabletypesize{\tiny}
%\tablewidth{\textwidth}
\resizebox{1.0\textwidth}{!}{
\begin{tabular}{lcccccccccc}
\hline \hline
ID & $\frac{1}{p} \frac{\diff p}{\diff t} \,  (\yr^{-1})$ & $\beta \,  (\mathrm{d \ cycle}^{-1})$ & $e$ & $f{(m)} \, (\Msun)$ & $a \sin i$ (AU) & $A$ (d) & $\omega$ (deg) & $\Porb$ (d) & Type & Refs. \\
\hline
	BL Cam & $-2.39(8) \times 10^{-8}$ & $-1.00(7) \times 10^{-13}$ & 0.8(7) & --- & 0.21(8) & --- & 81.29(11) & 5117.0 & SXP & \citet{Zong2019} \\
	DY Peg & $-(5.87 \pm 0.03) \times 10^{-8}$ & $(-8.55 \pm 0.04) \times 10^{-13}$ & $0.244 \pm 0.008$ & $(2.47 \pm 0.12) \times 10^{-5}$ & $0.353 \pm 0.005$ & $0.00204 \pm 0.00003$ & $239.3 \pm 5.2$ & $15425.0 \pm 205.7$ & SXP & \citet{Xue2020} \\
	KZ Hya & $(1.7 \pm 0.2) \times 10^{-7}$ & $(0.17 \pm 0.02) \times 10^{-11}$ & $0.25 \pm 0.01$ & $0.176 \pm 0.008$ & $5.02 \pm 0.010$ & 0.029  & --- & 9788.5 & SXP & \citet{Fu2008} \\
	\multirow{2}{*}{CY Aqr$^{a}$} & \multirow{2}{*}{$(2.132 \pm 0.002) \times 10^{-8}$} & \multirow{2}{*}{$(2.175 \pm 0.002) \times 10^{-13}$} & $0.139 \pm 0.002$ & $0.00063 \pm 0.00003$ & $1.32 \pm 0.02$ & $0.0076 \pm 0.0001$ & $-249.0 \pm 1.6$ & $21990.6 \pm 62.9$ & \multirow{2}{*}{SXP} & \multirow{2}{*}{This work}\\
	& & & $0.431 \pm 0.004$ & $0.00190 \pm 0.00006$ & $1.70 \pm 0.02$ & $0.0098 \pm 0.0001$ & $24.5 \pm 0.2$ & $18547.0 \pm 25.5$ & &\\
	DW Psc & --- & --- &  $0.40 \pm 0.05$ & $0.0405 \pm 0.0073$ & --- & $0.0066 \pm 0.0002$ & $257.5 \pm 5.1$ & 2220.7 & SXP & \citet{Qian2015}  \\
	V524 And & --- & --- &  $0.818 \pm 0.002$ & $(3.733 \pm 0.015) \times 10^{-5}$ & $0.1154 \pm 0.0604$ & $0.000433 \pm 0.000227$ & $158.2 \pm 44.6$ & 2342.3 & SXP & \citet{alton2019} \\
\hline
\end{tabular}
}
\begin{tablenotes}
\footnotesize{
\parbox[t]{\textwidth}{
			\item Notes: The types of primary stars are denoted as follows: SXP: SX Phoenicis variable; HADS: high amplitude $\delta$ Scuti variable; DSCT: $\delta$ Scuti variable. RRL: RR Lyrae variable. DCEP: classical Cepheids, or $\delta$ Cephei-type variables.
			\item $^a$: Variable stars with two companions.
			}} 
\end{tablenotes}
\end{center}
\end{table*}

\clearpage
\section{Long Tables} 
\label{app:TML}
%\clearpage
All the TML used in this work are listed in Table \ref{tab:TML}.

\startlongtable
\begin{deluxetable*}{cccc|cccc|cccc}
\label{tab:TML}
	\tablecaption{Times of maximum light of CY Aqr.}
	\tablehead{
		\colhead{TML($\BJD$)} & \colhead{Error} & \colhead{Detector} & \colhead{Source} & \colhead{TML($\BJD$)} & \colhead{Error} & \colhead{Detector} & \colhead{Source} &\colhead{TML($\BJD$)} & \colhead{Error} & \colhead{Detector} & \colhead{Source} 
	} 
	\startdata
		2427013.2934  & 0.0003  & pg    & 1     & 2454749.3874  & 0.0001 & CCD   & 1     & 2459833.76395  & 0.00003  & CCD   & 5 \\
		2427413.2204  & 0.0003  & pg    & 1     & 2454757.2612  & 0.0001 & CCD   & 1     & 2459833.82547  & 0.00003  & CCD   & 5 \\
		2427659.5071  & 0.0003  & pg    & 1     & 2454757.3221  & 0.0001 & CCD   & 1     & 2459833.88587  & 0.00003  & CCD   & 5 \\
		2427668.9734  & 0.0003  & pg    & 1     & 2454757.3832  & 0.0001 & CCD   & 1     & 2459833.94711  & 0.00002  & CCD   & 5 \\
		2427671.5303  & 0.0003  & pg    & 1     & 2454761.2899  & 0.0001 & CCD   & 1     & 2459834.00803  & 0.00004  & CCD   & 5 \\
		2427671.5321  & 0.0003  & pg    & 1     & 2454761.3510  & 0.0001 & CCD   & 1     & 2459834.06907  & 0.00003  & CCD   & 5 \\
		2427682.5215  & 0.0003  & pg    & 1     & 2454781.2492  & 0.0001 & CCD   & 1     & 2459834.13050  & 0.00003  & CCD   & 5 \\
		2427684.9633  & 0.0003  & pg    & 1     & 2454781.3098  & 0.0001 & CCD   & 1     & 2459834.19107  & 0.00004  & CCD   & 5 \\
\enddata
	\tablecomments{The numbers in the fourth column indicate the different data sources:
	1: \citet{fang2016}; 2: \citet{Wiedemair2018}; 3: \citet{Wiedemair2020}; 4: AAVSO; 5: TESS.		
	}
\end{deluxetable*}

\end{CJK*}
\end{document}